\documentstyle[graphicx,multicol,prl,aps]{revtex}

\newcommand{\myscalebox}[1]{\scalebox{0.4}[0.4]{#1}}

\begin{document}
\pagestyle{empty}
\begin{multicols}{2}
\noindent

{\bf Comment on ``Glassy Transition in a Disordered Model for the RNA
  Secondary Structure''}

In a recent, very interesting paper, Pagnani, Parisi and Ricci-Tersenghi 
\cite{pagnani2000} have studied the low-temperature behavior of a model
for RNA secondary structure. They claim that the model exhibits a
breaking of the replica symmetry, since the width of the
distribution $P(q)$ of overlaps may converge to a finite value at
$T=0$. The authors used an exact enumeration method to obtain all
ground states for a given RNA sequence. Because of the exponential
growing degeneracy, only sequences up
to length $L=256$ could be studied.

Here it is shown that, in contrast to the previous results,
by going to much larger sizes as $L=2000$ 
the variance $\sigma^2(q) \propto L^{-0.5}$. 
This means that $P(q)$ becomes 
a delta function in the thermodynamical limit at $T=0$.
The method used here combines the ideas presented in
\cite{pagnani2000} and \cite{higgs1996}. The method is faster than the
algorithm of \cite{higgs1996} because no floating-point arithmetic is
necessary. Furthermore, the algorithm of \cite{higgs1996} is not exact,
although usually true ground states are obtained. The technique of
\cite{pagnani2000} guarantees exact ground states but is restricted to
small sizes.

Here, a finite number of exact ground states is
selected randomly from the set of all ground states which is
represented by a graph. Similar to an ordinary Monte-Carlo simulation it
has to be guaranteed that each ground state appears with the proper
weight, i.e. with the same probability, since all ground states have
exactly the same energy. This is ensured by the following technique:
Let now $G_{i,j}$ denote the set of ground states for the sequence
$[r_i,\ldots,r_j]$. 
 Similar to the representation of the partition function
 applied in \cite{higgs1996,pagnani2000},  
$G_{i,j}$ can be expressed in
 terms of ground states for smaller sequences:
a ground state
for the sequence $[r_i,\ldots,r_j]$ can either be a ground-state of
$[r_{i+1},\ldots,r_{j}]$ (if the energy is low enough), or it is a
combination of a pair $(i,k)$ ($k\in (i+1,\ldots,j)$) with an
arbitrary ground state of $[r_i,\ldots,r_{k-1}]$ and an arbitrary
ground state of $[r_{k+1},\ldots,r_j]$ (if the energy is low enough).

The calculation of all ground states proceeds as follows:
$G_{i,i}=G_{i,i+1}=\emptyset$ for all feasible $i$ holds. 
Starting with $G_{i,i+2}=\{(i,i+2)\}$ ($i=1,\ldots,L-2$) the
complete set of ground states can be calculated recursively. The
result is stored as a acyclic 
directed graph with $G_{i,j}$ ($1\le i\le j\le L$)
being the nodes and $G_{1,L}$ the root. At each
node, edges pointing to to the descendant sets $G_{i+1,j}$, $G_{i+1,k-1}$ and 
$G_{k+1,1}$ are stored instead of enumerating the states. Additionally,
along with each node the ground-state energy $E_0(i,j)$ and the number of
ground states $d_{i,j}$ is kept. The degeneracy $d_{i,j}$ can be
calculated recursively as well. 
 

The selection of a ground state is performed by a steepest descent
into the graph. Each ground state consists of the pairs encountered during
the descent. At each node the steepest descent continues either into
one
descendant $G_{i+1,j}$ or into two descendant $G_{i+1,k-1}, G_{k+1,j}$,
the alternative for proceeding being chosen randomly. The
probability for each choice is proportional to the number of ground
states found in the corresponding branch(s). For that purpose
 the degeneracy values
$d_{i,j}$ are used. It means that a path which contains twice the
number of ground states of another path is selected on average twice as
often. Therefore, it is guaranteed that
each single ground state contributes with the same weight and a
statistical correct $T=0$ average is obtained.

For each sequence length the calculations were
performed for 8000 independent
realizations of the disordered, except for $L=2000$ where only 1800
random sequences were generated.  For each realization 100 ground
states were selected randomly and stored for further evaluation. It
was tested that by increasing this number the results do not change
significantly. 

\begin{figure}[ht]
\begin{center}
\myscalebox{\includegraphics{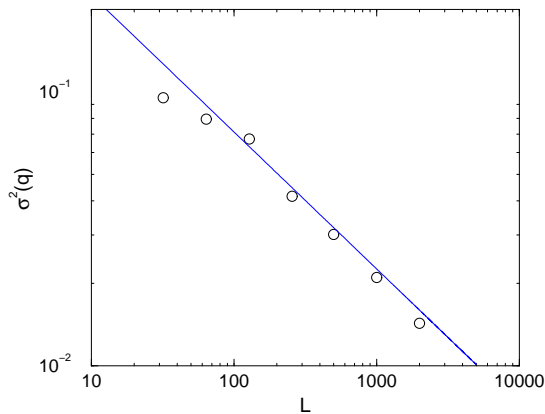}}
\end{center}
\caption{Variance
 $\sigma^2(q)$ of the distribution of overlaps as a function of
  system size $L$. The line represents the function $0.71L^{-0.5}$.
Please note the double logarithmic scale.
}
\label{figSigma}
\end{figure}

The resulting values $\sigma^2(q)$ are
shown in Fig. \ref{figSigma} using a double logarithmic
scale. Clearly, the function converges towards zero, thus $P(q)\to
\delta(q)$ for $L \to\infty$, a similar result was found
\cite{alex-rna} for the model
presented in \cite{higgs1996}. For small sizes,
this convergence is much smaller due to finite-size effects. This
may be the reason that in \cite{pagnani2000} no decision about the
behavior of the width of $P(q)$ could be taken.

The author thanks A. Pagnani and F. Ricci-Tersenghi for interesting
discussions on the subject.

\noindent 
A.K. Hartmann\\
{\small   Institut f\"ur theoretische Physik,
G\"ottingen, Germany}

\end{multicols}

\begin{references}

\bibitem{pagnani2000} 
  A. Pagnani, G. Parisi and F. Ricci-Tersenghi,
  Phys. Rev. Lett. {\bf 84}, 2026 (2000)

\bibitem{higgs1996} P.G. Higgs, Phys. Rev. Lett {\bf 76} (1996)

\bibitem{alex-rna} A.K. Hartmann, unpublished
\end{references}
\end{document}